\documentclass[preprint,3p,times,twocolumn]{elsarticle}

\begin{document}
  \begin{frontmatter}

    \title{The 4\,K outer cryostat for the CUORE experiment:\\ construction and quality control}
    
    \author[a]{F.~Alessandria}
    \author[b,c]{M.~Biassoni}
    \author[b]{G.~Ceruti}
    \author[d]{A.~Chiarini}
    \author[b,c]{M.~Clemenza}
    \author[b]{O.~Cremonesi}
    \author[e]{V.~Datskov}
    \author[b]{S.~Dossena}
    \author[b,c]{M.~Faverzani}
    \author[b,c]{E.~Ferri}
    \author[b,c]{A.~Nucciotti}
    \author[b]{M.~Perego}
    \author[b]{E.~Previtali}
    \author[b,c]{M.~Sisti}
    \author[f]{L.~Taffarello}
    \author[b,c]{F.~Terranova}
    
    \address[a]{INFN Sez. di Milano, Milan, Italy}
    \address[b]{INFN Sez. di Milano-Bicocca, Milan, Italy}
    \address[c]{Dipartimento di Fisica, Universit\'a di Milano-Bicocca, Milan, Italy}
    \address[d]{INFN Sez. di Bologna, Bologna, Italy}
    \address[e]{CERN, Geneva, Switzerland}
    \address[f]{INFN Sez. di Padova, Padova, Italy}
      
      \begin{abstract}
The external shell of the CUORE cryostat is a large cryogen-free
system designed to host the dilution refrigerator and the bolometers
of the CUORE experiment in a low radioactivity environment. The three
vessels that form the outer shell were produced and delivered to
the Gran Sasso underground Laboratories in July 2012. In this paper,
we describe the production techniques and the validation tests done at 
the production site in 2012.
      \end{abstract}
      
  \end{frontmatter}
  

  \section{Introduction}

\begin{figure}[ht]
\centering
\includegraphics[width=0.95\linewidth,type=pdf,ext=.pdf,read=.pdf]{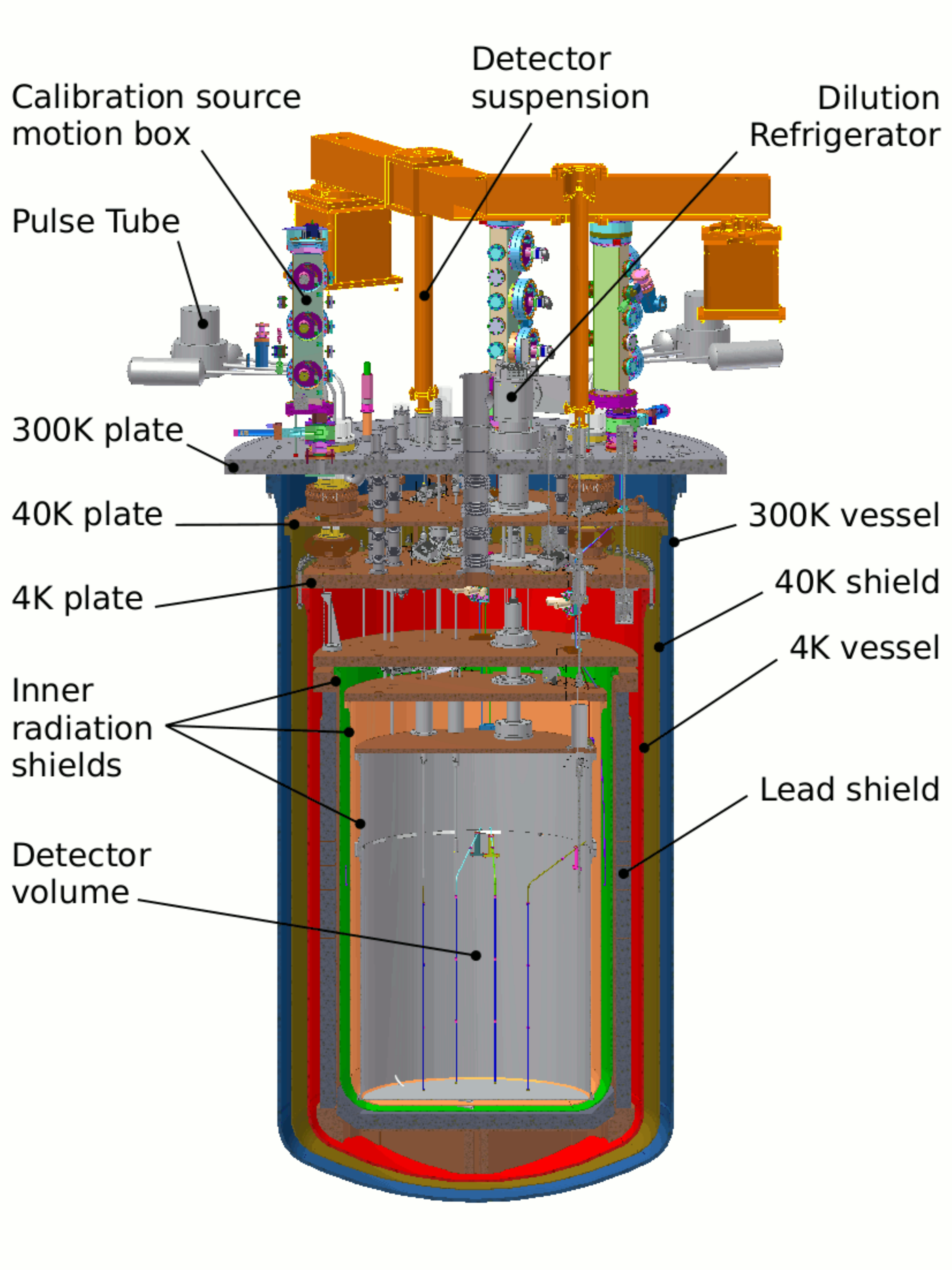} 
\caption{The CUORE cryostat.}
\label{fig:cryostat}
\end{figure}

CUORE \cite{Arnaboldi:2003tu,Ardito:2005ar,Alessandria:2011rc} is a
ton-scale neutrinoless double beta decay experiment currently under
construction in the Gran Sasso Underground Laboratory (Laboratori
Nazionali del Gran Sasso, LNGS, Italy). The detector will consist of
988, $5 \times 5 \times 5$~cm$^3$ TeO$_2$ bolometers operating at very
low temperature ($T \approx 10$~mK). The active mass of CUORE (740~kg)
exceeds the mass of its precursor (40.7~kg~\cite{Andreotti:2010vj}) by
more than one order of magnitude and sets unprecedented challenges
both in terms of material radiopurity and cooling power.  The CUORE
cryostat~\cite{cryostat,cryostat2} must deliver appropriate cooling
power to keep the bolometers and their support structure (about
1500~kg overall mass) at 10~mK base temperature. In addition, the
cryostat operation must be reliable enough to guarantee year-long data
taking and must not introduce microphonic noise, which translates into
instabilities in the detector readout. Finally, the materials have to
be carefully chosen in order to reach background levels
$\le10^{-2}$~counts/keV$\cdot$kg$\cdot$y~\cite{Alessandria:2011vj}.
The design of CUORE has taken advantage of the progresses in
cryogen-free refrigerators over the last twenty
years~\cite{review_cryogenfree}.  The CUORE cryostat is made up of two
vacuum tight vessels and four thermal shields
(Fig.~\ref{fig:cryostat}). The two outer vessels and the shield
between them constitute the outer cryostat which will maintain a 4\,K
environment (``4\,K cryostat'') by means of two stage Pulse Tube
refrigerators~\cite{mikulin} which will cool the inner vacuum tight
vessel to about 4\,K; the inner thermal shields will be cooled by the
thermal stages of a He$^3$/He$^4$ Dilution refrigerator as well as the
detector.  Radiopurity requirements for the 4\,K cryostat are less
demanding since the inner volume is screened by a low activity lead
shield. Nevertheless, the use of steel cannot be afforded and special
care must be put in the welding procedures to avoid presence of
contaminants. In this paper we describe the design of the outer 4\,K
cryostat and its requirements to operate in the cryogenic apparatus of
the CUORE experiment (Sec.~\ref{sec:outer_shell}). 
Sec.~\ref{sec:production} describes the fabrication of the 300\,K and 4\,K
vessels, the 40\,K thermal shield and
the technical choices employed to fulfill the above requirements.  The
cleaning procedure is detailed in Sec.\ref{sec:cleaning}.  Finally,
the validation tests done at the production site are described in
Sec.\,\ref{sec:test_setup} and \ref{sec:results}.

\section{The 4\,K cryostat: design and specifications}
\label{sec:outer_shell}

The CUORE cryostat consists of six nested vessels and shields. 
The three outermost ones are included in the outer Cryostat. 
Two of them (300\,K and 4\,K) are vacuum
tight: the space between the 300\,K and 4\,K vessels constitutes the
Outer Vacuum Chamber (OVC) of the cryostat and the 300\,K vessel
operates at room temperature.  The volume inside the 4\,K vessel
represents the Inner Vacuum Chamber (IVC) and in normal running
condition this vessel is thermalized to 4\,K. Between the 300\,K and
4\,K vessel there is a thermal radiation shield at 40\,K covered with
multi-layer aluminized superinsulation. Up to five pulse tubes (PT) mounted on the OVC top
plate provide the cooling for the 40\,K radiation shield (PT first
stages) and the IVC (PT second stages).  The IVC encloses the
detector and the shielding lead with a volume of about 3~m$^3$. 

The mass to be cooled and the tight requirements on the bulk
radioactivity strongly constraint both the material choices and the
production techniques that can be employed for the outer cryostat. In
fact, in order to preserve radiopurity the use of copper for the
vessels and shields is mandatory and therefore stainless stainless
steel cannot be used. Most conventional
welding techniques cannot be employed since potentially they can increase
the bulk radioactivity due to filler materials and electrodes. In
addition, the 4\,K cryostat will be operated both in PT-cooling mode
and in fast cooling mode. When cooling power is provided just by the
pulse tubes (PT-cooling mode), both the IVC and OVC must be under
vacuum ($<10^{-4}$~mbar at room temperature). However, to speed up the
cooling of the shielding lead (fast cooling mode), cold Helium gas will be
injected into the IVC.  For proper operation, the Helium in the IVC
will be maintained at an absolute pressure of about 130 kPa (1.30 bara).  Hence,
the mechanical properties and vacuum tightness must be sustained by the
IVC and OVC separately and under overpressure conditions.

The most relevant specifications for
the 4\,K outer shell can be summarized as follows:

\begin{itemize}
\item {\it Vacuum} Both the 300~K and 4~K vessels must be able to stand vacuum
at room temperature ($p<10^{-4}$~mbar); the vacuum tightness of the weldings
and of the seals (elastomer orings for the OVC, indium and metallic gasket for the IVC)
must be kept at the $10^{-9}\mathrm{mbar}\cdot\mathrm{l/s}$ level
\item {\it Pressure} The 4~K and 300~K vessels must be able to stand
both internal and external overpressure according to the design
specification of Tab.~\ref{tab:pressure} without mechanical 
deformations or worsening of the above-mentioned vacuum
specifications
\item {\it Thermal properties} The vessels must be able to preserve vacuum and pressure specifications
after multiple thermal cycles down to 4~K
\item {\it Radiopurity} The $^{232}$Th and $^{238}$U contamination of all cryostat thermal shields/vessels
must be $<2 \times 10^{-6}$~Bq/kg  and $<10^{-4}$~Bq/kg, respectively.  
  
\end{itemize}

The vacuum and pressure specifications were tested in a direct manner
at the production site and are described in Sec.\,\ref{sec:test_setup}
and \ref{sec:results}. In addition, since thermal contraction effects
saturate already at liquid Nitrogen temperature (77~K), a dedicated
colt test at 77~K was also done, as described in
\ref{sec:cold_tests}. The bulk radiopurity of the copper selected for
the production of the 4~K cryostat (see Sec.~\ref{sec:production}) was
sampled during the material procurement. No additional radiopurity
tests were performed at the production site and the contributions due
to machining and surface cleaning will be determined during the
commissioning runs at LNGS.

\begin{center}
\begin{table}[ht]
{\small
\hfill{}
\begin{tabular}{lll}
\hline
&\multicolumn{2}{c}{$\Delta P_{max}$ [bar]}\\
vessel & internal & external \\
\hline
OVC & 0.1 & 1.01 \\
IVC & 1.5 & 1.1 \\
\hline
\end{tabular}}
\hfill{}
\caption{Design operating pressure for OVC and IVC vessels.}
\label{tab:pressure}
\end{table}
\end{center}

\section{Production of the 4\,K cryostat}
\label{sec:production}

The vessel shells and torispherical heads are made of selected high purity
copper produced by Norddeutsche Affinerie~\cite{aurubis}; the chosen
copper alloy for the outer cryostat is the Oxygen Free Electrolytic
(OFE) one. Assays of Cu-OFE samples set only an upper limit of
about $2\times10^{-11}$\,g/g for the $^{232}$Th content, which is
appropriate for the CUORE background requirements mentioned in
Sec.~\ref{sec:outer_shell}.  The 300\,K vessel is also made of Cu-OFE,
except for the upper flange, which is build from austenitic stainless
steel (EN 1.4307, AISI 304L). Similarly, the top plates for the 40\,K
shield and 4\,K vessel are copper-made, while the top plate of the
300\,K vessel is made of stainless steel. The mechanical properties of
the copper and stainless steel
alloys employed are listed in Table~\ref{tab:materials}: the Cu-OFE
R220 copper alloys have been used exclusively for the torispherical
heads to ease the forming. The use of stainless steel for the top
flange and plate of the 300\,K vessel guarantees a better mechanical
stability and vacuum tightness. The same choice was not possible for
the 4\,K vessel because of the poor thermal conductance of stainless
steel.  Table~\ref{tab:vessels} summarizes the main mechanical
parameters of the outer cryostat. The wall thicknesses were determined
by the operating pressure requirements (Table~\ref{tab:pressure}) and
were calculated according to the ASME\textcopyright\ code
\cite{asme}. A linear buckling analysis performed with
ANSYS~\cite{ansys} gives a limiting differential pressure of 2.0 and
2.3 bar for the OVC and IVC vessel, respectively.  For what concerns
the top 300\,K and 4\,K plates, dimensions were chosen accounting for
the static mechanical loads (vessels and lead shield) and possible
additional loads induced by seismic events.
\begin{center}
\begin{table*}[ht]
{\small
\hfill{}
\begin{tabular}{llll}
\hline
 & Ultimate Strength & Yield Strength & Elongation \\
Material & $Rm$ [N/mm$^2$]& $Rp0,2$ [N/mm$^2$]& $A5$ [\%]\\
\hline
1.4306 304L & 520-670 & 200 \\
C10100 (Cu-OFE) R240 & 240-300 & $>180$ & $>15$ \\
C10100 (Cu-OFE) R220 & 220-260 & $>100$ & $>30$ \\
\hline
\end{tabular}}
\hfill{}
\caption{Mechanical properties of the steel and OFE copper.}
\label{tab:materials}
\end{table*}
\end{center}
\begin{center}
\begin{table*}[ht]
{\small
\hfill{}
\begin{tabular}{llllll}
\hline
 & Inner diam. (mm) & Height (mm) & Thickness (mm) & Weight (kg) & Material \\
\hline
300\,K vessel & 1603 & 3030 & 12 & 1942 & Cu-OFE+304L \\
40\,K vessel & 1503 & 2765 & 5 & 681 & Cu-OFE \\
4\,K vessel & 1363 & 2471 & 10 & 1142 & Cu-OFE \\ 
300\,K plate & 2060 & - & 63 & 1532 & 304L \\
40\,K plate & 1573 & - & 20 & 308 & Cu-OFE \\
4\,K plate & 1473 & - & 59 & 870 & Cu-OFE \\
\hline
\end{tabular}}
\hfill{}
\caption{Main mechanical parameters of the 4\,K cryostat components.}
\label{tab:vessels}
\end{table*}
\end{center}

\begin{figure}[ht]
\centering
\includegraphics[scale=0.3,type=pdf,ext=.pdf,read=.pdf]{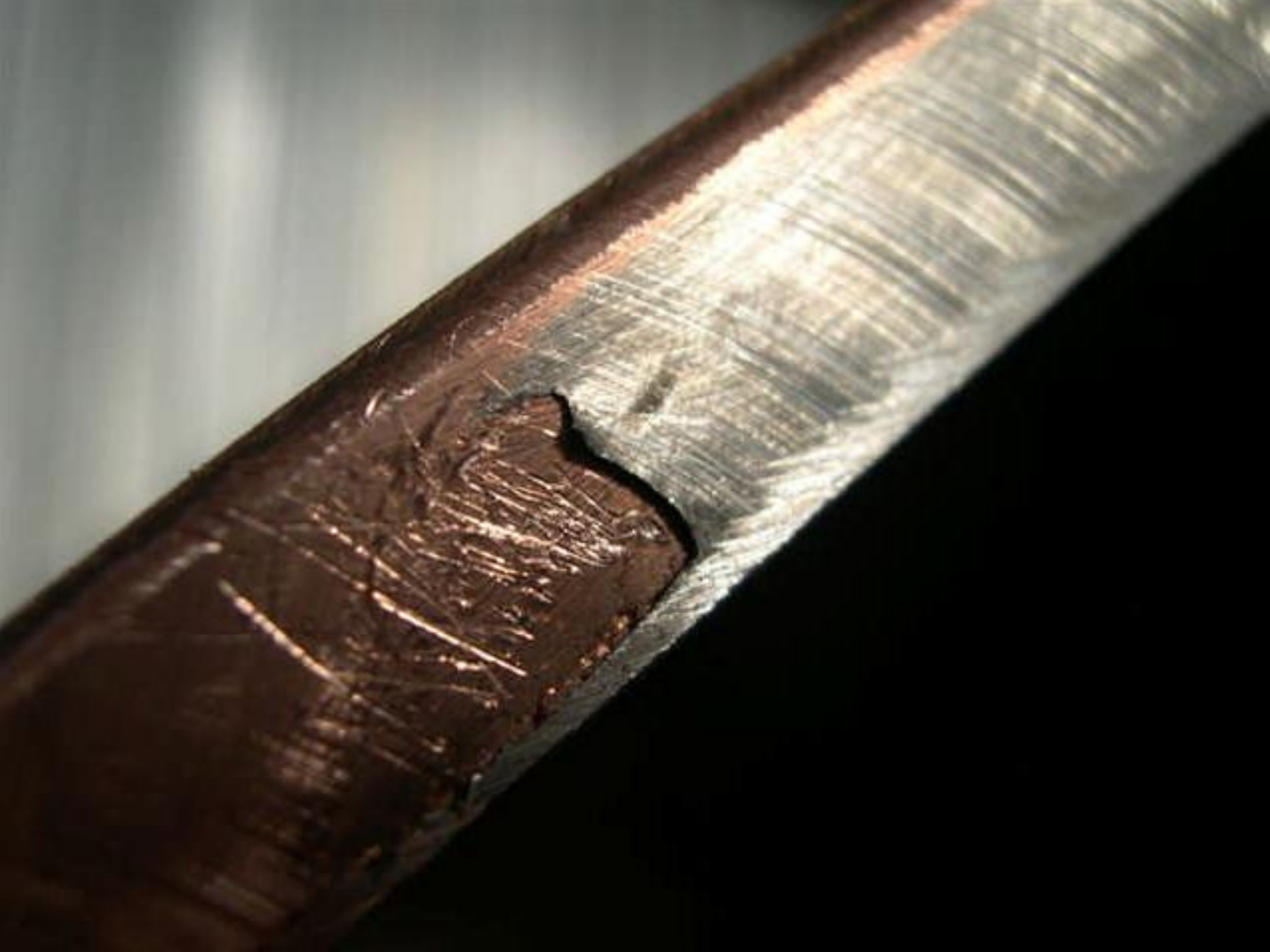} 
\caption{Defective welding at the steel-copper interface of the 300\,K vessel.}
\label{fig:lack_of_fusion}
\end{figure}

The vessels, shields, and plates were produced, ca\-len\-de\-red and
machined by Simic s.p.a.~\cite{simic} in 2010-2012.  In order to
minimize radioactive contamination, the copper plates were joined by
electron beam welding (EBW). In fact EBW is a contactless welding
technique which does not require either filler materials or electrodes
and which minimizes the heat affected zone~\cite{ekin,shultz}.  It is
also particularly suited for vacuum applications; still, EBW in CUORE
required special care since there is little experience in welding
thick copper plates.  All copper-copper weldings were carried out by
Pro-beam GmbH~\cite{probeam} employing the EBW technique. The
steel-copper interface for the 300\,K vessel was welded by Pro-beam
GmbH through EBW, too.  Vessel and shield shells are fabricated
starting from two copper plates which are first calendered and
longitudinally welded.  The cylinder pairs are then joined together,
to the heads, and to the top flanges by three radial weldings.  The
quality of all welding seams was assessed at Pro-beam by means of dye
penetration tests and X-rays.  All weldings were also individually
tested for helium leaks with a specially designed tool at Simic.  Both
type of weldings required several iterations to be tuned.  In
particular, the steel-copper weld turned out to be quite challenging
mostly due to an uncompensated deflection of the electron beam in the
proximity of the Cu-304L interface that resulted in a significant lack
of fusion during the first welding attempt
(Fig.~\ref{fig:lack_of_fusion}). The problem was solved using an
inclined beam ($5^\circ$ at a distance of 500~mm) and re-tuning the
electron accelerating parameters of the Pro-beam machines. In order to
achieve a larger margin of safety, the EBW was done joining the copper
vessel with a 10~cm high steel ring. The steel flange of the 300\,K
vessel was joined to the ring later on, using standard Gas Tungsten
Arc welding (GTAW).  All GTAW weldings were performed using radiopure
tungsten electrodes; they were selected by the INFN Radioactivity
Laboratory located at University of Milano-Bicocca.  Some of the
Cu-Cu welds of the 300\,K and 4\,K vessel showed a significant level of
porosity, which required two additional EBW passages. A second EBW
passage was applied also to the steel-copper weld.  The final quality
assessment of the welds and, in particular, X-ray checks excluded
lack of fusion. However, shallow and mid-depth porosities were still
visible (Fig.\ref{fig:porosities}). Direct vacuum and pressure tests
(see Sec.~\ref{sec:results} below) demonstrated that these porosities
do not weaken neither the vacuum tightness nor the structural
properties of the vessel.

\begin{figure}[ht]
\centering
\includegraphics[scale=0.3,type=pdf,ext=.pdf,read=.pdf]{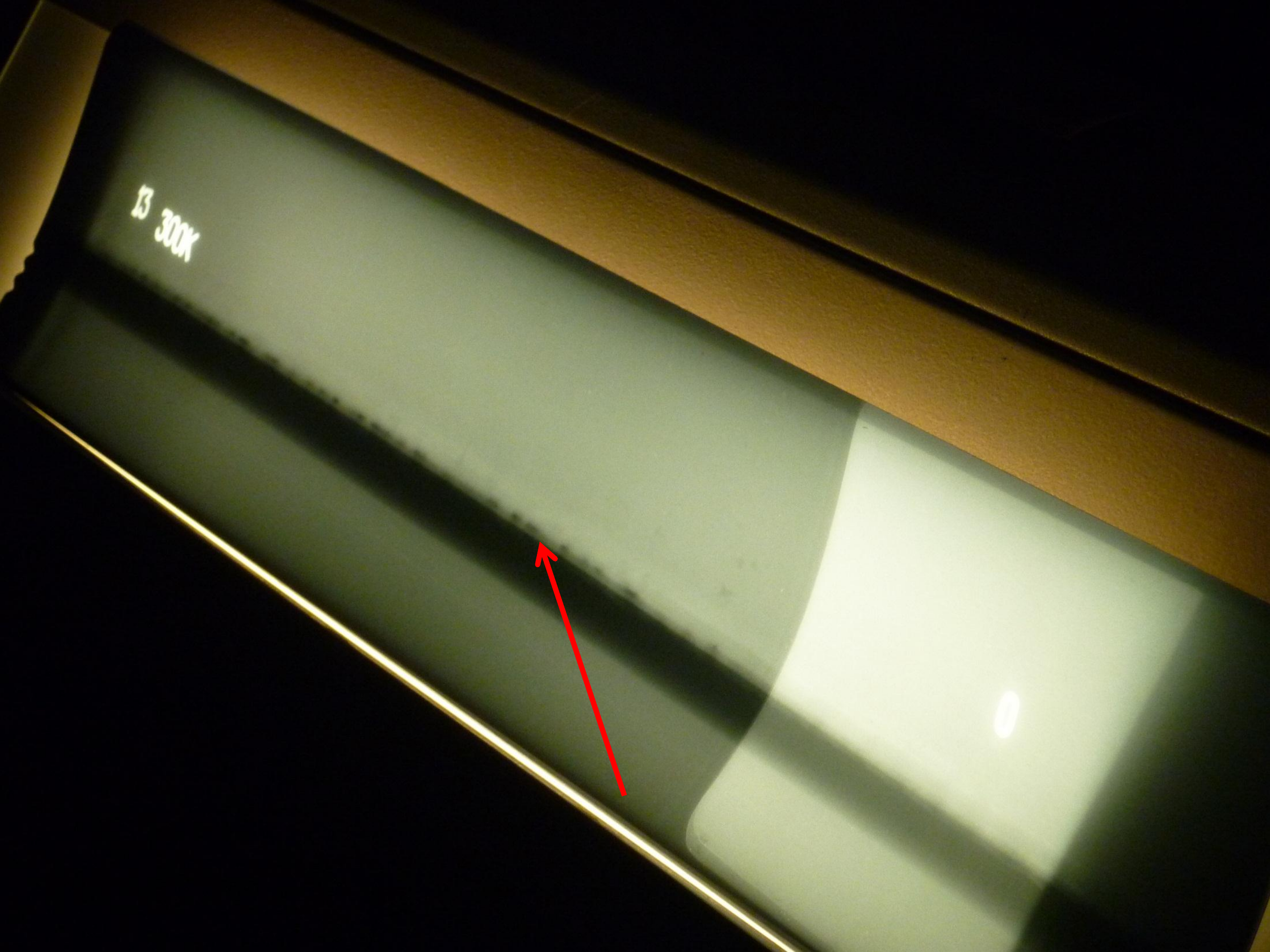} 
\caption{X-ray mark of the steel-copper welding. The gray spots pointed by the arrow
are indication of shallow and mid-depth porosities.}
\label{fig:porosities}
\end{figure}

\begin{figure}[ht]
\centering
 \includegraphics[width=0.5\textwidth,keepaspectratio=true]{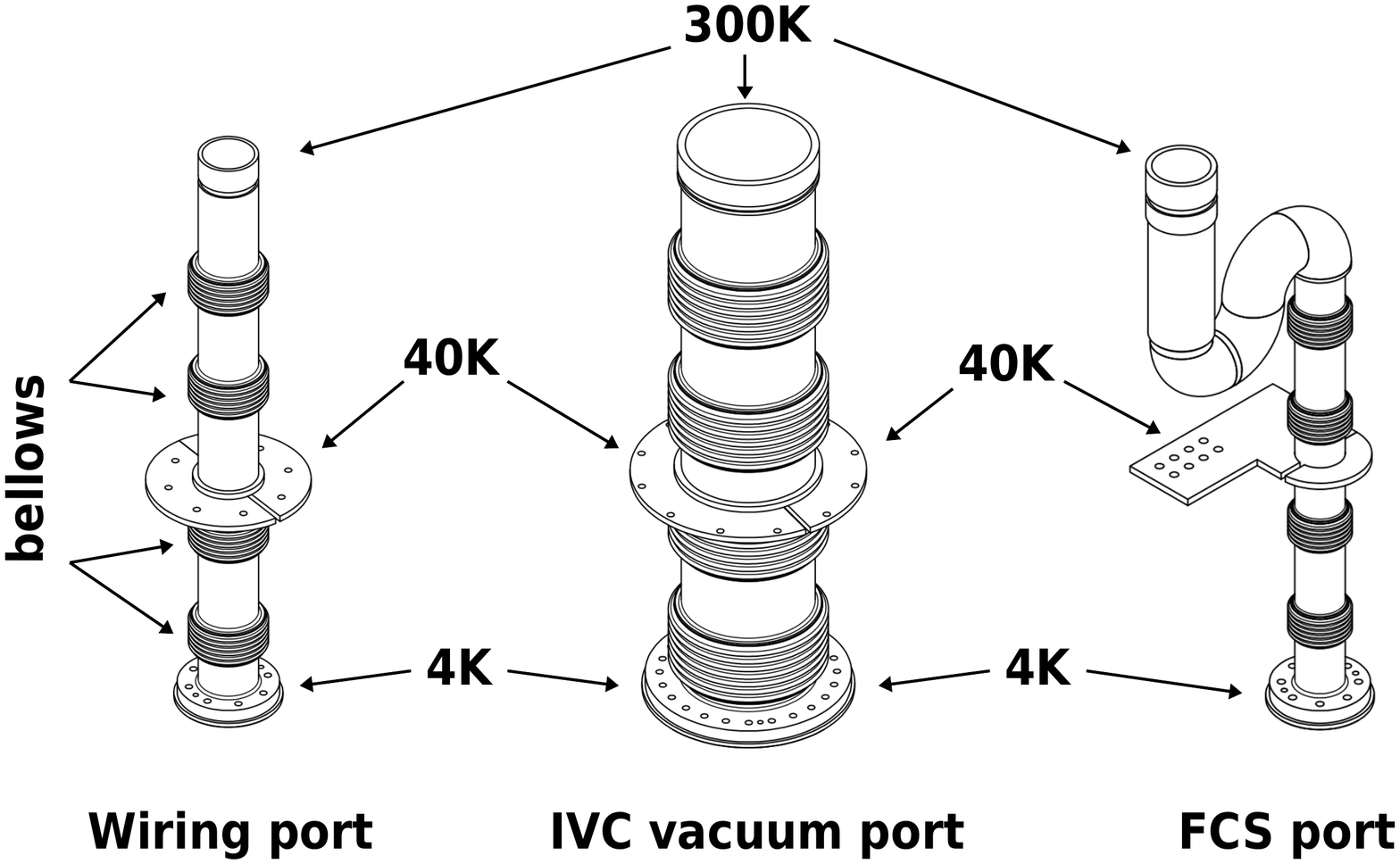}
\caption{The wiring port, the IVC vacuum pumping port and the
  Fast Cooling System port.}
\label{fig:ports}
\end{figure}

\begin{figure}[ht]
\centering
\includegraphics[scale=0.31,type=pdf,ext=.pdf,read=.pdf]{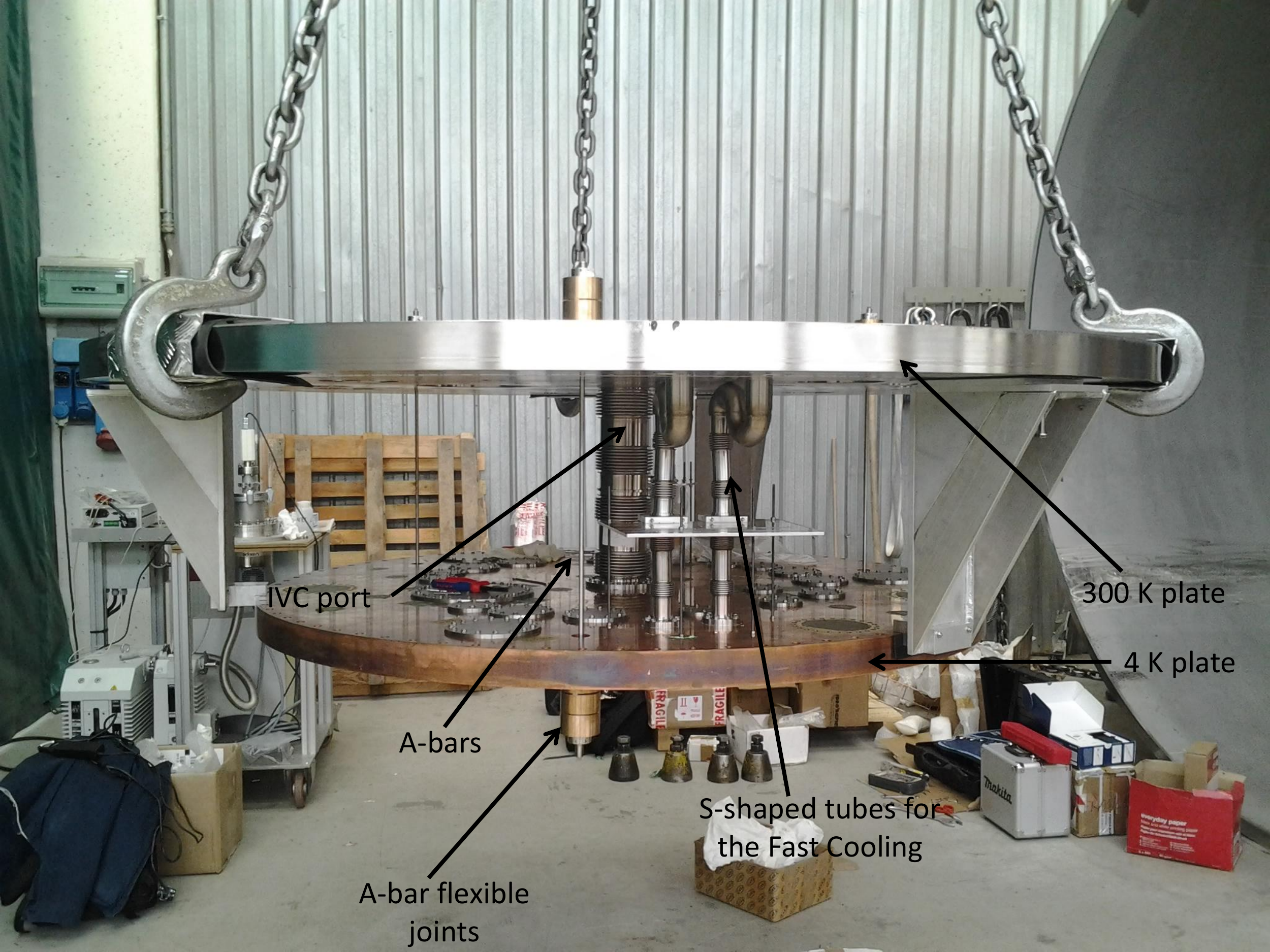} 
\caption{The 300\,K and 4\,K plates before positioning into the
  vessels. The plates are connected by the A-bars and the
  fast cooling tubes, the IVC vacuum tube and (not visible in the photo) the
  wiring tube are installed.}
\label{fig:plates}
\end{figure}

The outer 4\,K cryostat components are held by the 300\,K plate,
which in turn is connected to the main support structure of the
experiment. The 300\,K plate holds the corresponding vessel by bolts
and an elastomer seal. The 40\,K and 4\,K plates are held below the 300\,K 
plate by three 8~mm diameter steel bars: the ``A-bars'' that link
the 300~K plate to the 4~K plate and the ``B-bars'', linking the 300~K
plate to the 40~K plate. The bars are made of EN 1.4429 (AISI 316LN) steel for cryogenic applications. 
Vacuum tightness of the 4~K vessel at the plate-vessel
interface is obtained employing a tubular metallic seal (Garlock
Helicoflex\textsuperscript{\textregistered}, model HNV 200) mounted inside a
centering ring (stainless steel 316~LN). The bars are linked to the plates by
tilting Cu-Be joints. Both the A-bars and the B-bars, when loaded, can be tilted
with respect to the vertical by at most $1.5^\circ$ without damage. 
This design allows the system to withstand seismic events with a peak ground
acceleration as large as about 0$.08 g$. Finally, several
feedthroughs will be employed in CUORE for pumping, fast cooling with
cold He and for the electrical leads of the detector (Fig.~\ref{fig:ports}). During the tests
at the production site we mounted only a few of them: the vacuum ports
for the IVC and OVC, the two ports of the fast cooling and one wiring
port. Fig.~\ref{fig:plates} shows the 300\,K and 4\,K plates connected
by the A-bars with the above-mentioned ports before closing the chambers
(see Sec.~\ref{sec:test_setup}).

  \section{Cleaning}
\label{sec:cleaning}

\begin{figure}[ht]
\centering
\includegraphics[scale=0.065,type=pdf,ext=.pdf,read=.pdf]{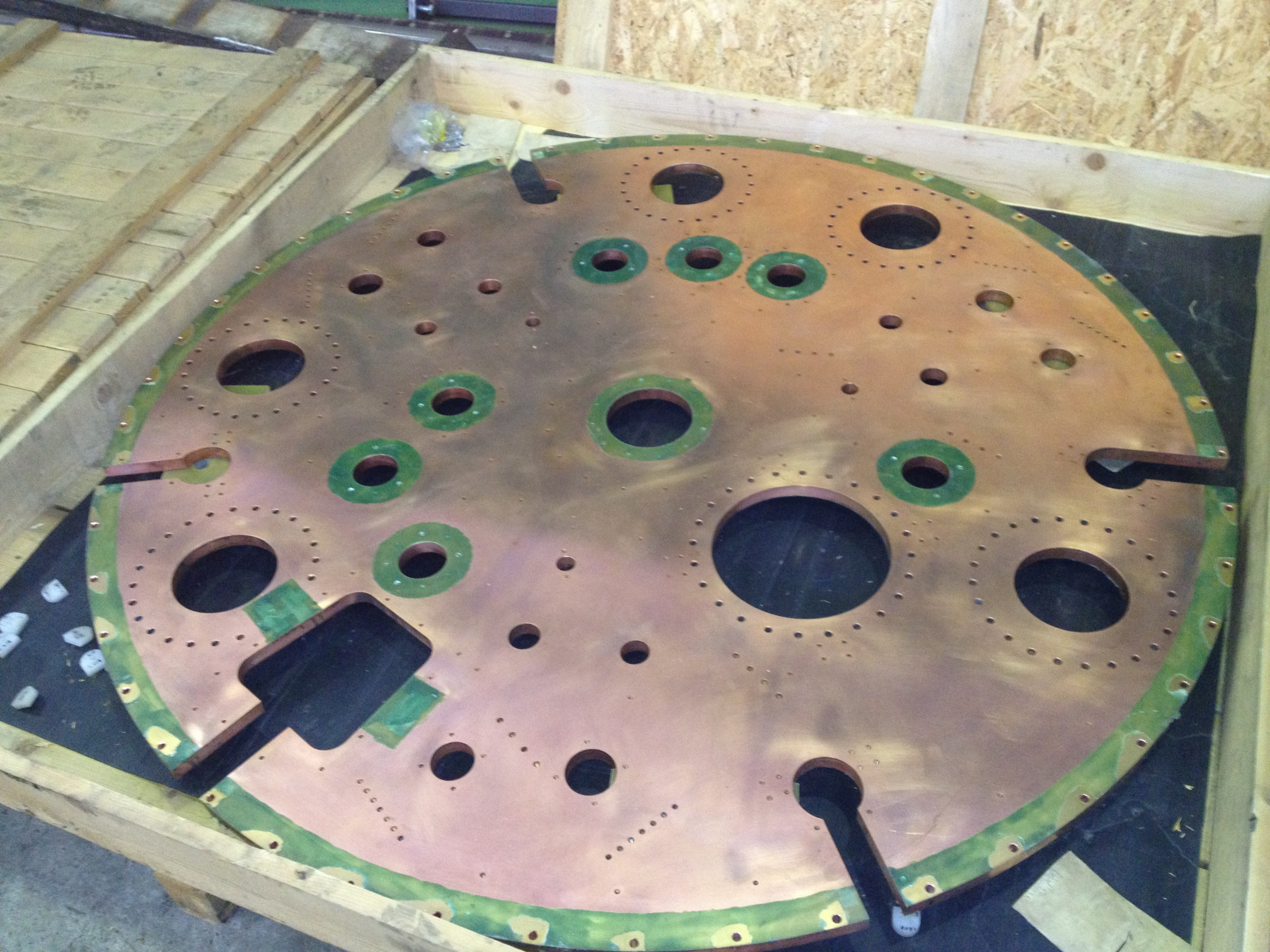} 
\caption{The 40~K plate with gilded flanges protected by a plastic layer.}
\label{fig:4Kplate}
\end{figure}

The production of the vessels from the pre-cut copper plates includes
calendering, welding and machining. Similarly, the plates were
machined and drilled, and several flanges were gold plated in order to
improve thermalization (Fig.~\ref{fig:4Kplate}). The copper has been
selected for its very low bulk radioactivity, therefore special care
has been put into the welding and milling procedures to preserve its
purity. However, surface tratment after machining is needed to remove
surface contaminations that can compromise radiopurity, vacuum
(surface degassing) and thermal (emissivity) properties.  In fact,
from the radioactivity point of view, the surface cleaning of the
cryostat outer shell is not particularly critical in CUORE since none
of these parts are directly facing the bolometers. Unlike the
innermost CUORE copper pieces~\cite{copper_cleaning} in contact to the
detector, here the removal of surface contaminants was achieved by
standard metal treatment for industrial applications. Still,
degreasers, acids and solvents used during the processing of the
metals were pre-selected for radiopurity. In particular, we have
applied the treatment of the copper successfully employed for
CUORICINO~\cite{Andreotti:2010vj}.

The cleaning of the vessels and copper plates were performed by BAMA
s.r.l.~\cite{bama} employing a procedure that can be summarized as follows:

\begin{itemize}
\item {\it Degreasing} The degreaser employed in this phase is a
  general purpose decontamination solution
  (Radiacwash\textsuperscript{\textregistered}) used in nuclear
  medicine for the removal of radioactive contaminants.  It was
  diluted (10\% in weight) using purified water with a conductivity
  $<5 \mu\mathrm{S} \cdot \mathrm{cm}^{-1}$.  The solution was sprayed
  on the vessels.
\item{\it Flushing} Performed with purified water (conductivity
  $<5 \mu\mathrm{S} \cdot \mathrm{cm}^{-1}$) at high pressure ($>$100 bar).
  Three to four degreasing and flushing cycles were performed per vessel.
\item{\it Pickling} Pickling was carried out by nitric acid in
  water solution (10\% weight). For all vessels, the
  degreasing and pickling procedure was performed twice.
\item{\it Passivation} All copper parts were treated with citric acid in water solution (10\% weight).
\item{\it Final flushing} The last flushing of each item was done by
  high purity water (conductivity $<0.5 \mu\mathrm{S} \cdot \mathrm{cm}^{-1}$,
  Grade 2 ISO 3696).
\item{\it Packing} All vessels and plates were packed in Nitrogen
  atmosphere using an aluminized wrapping. The wrapping has been tested for
  radiopurity before the start of the cleaning procedure.
\end{itemize}

\begin{figure}[ht]
\centering
\includegraphics[scale=0.38,type=pdf,ext=.pdf,read=.pdf]{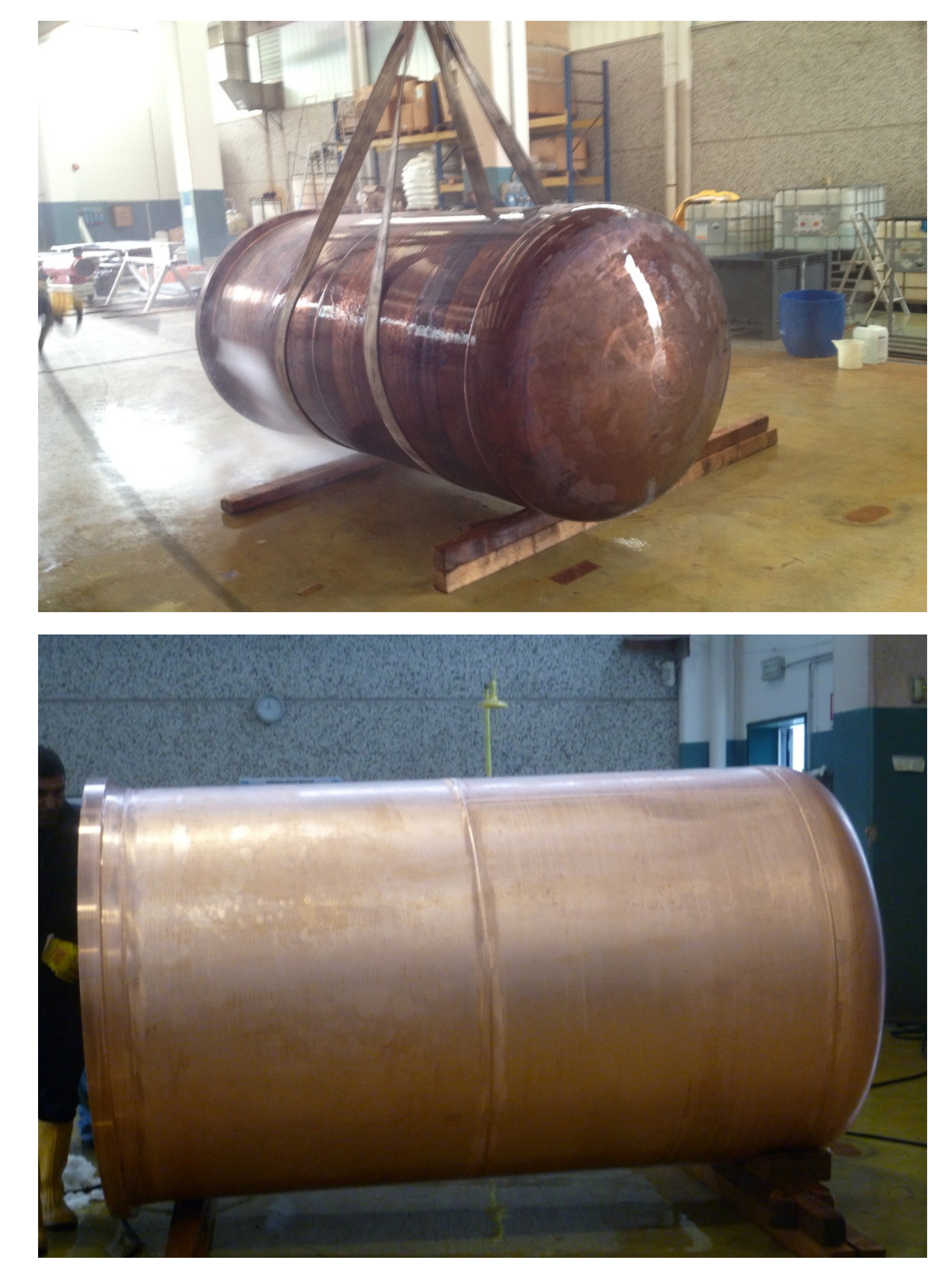} 
\caption{The 4~K vessel at the beginning and at the end of the cleaning procedure.}
\label{fig:cleaning_4k}
\end{figure}

\begin{figure}[ht]
\centering
\includegraphics[scale=0.3,type=pdf,ext=.pdf,read=.pdf]{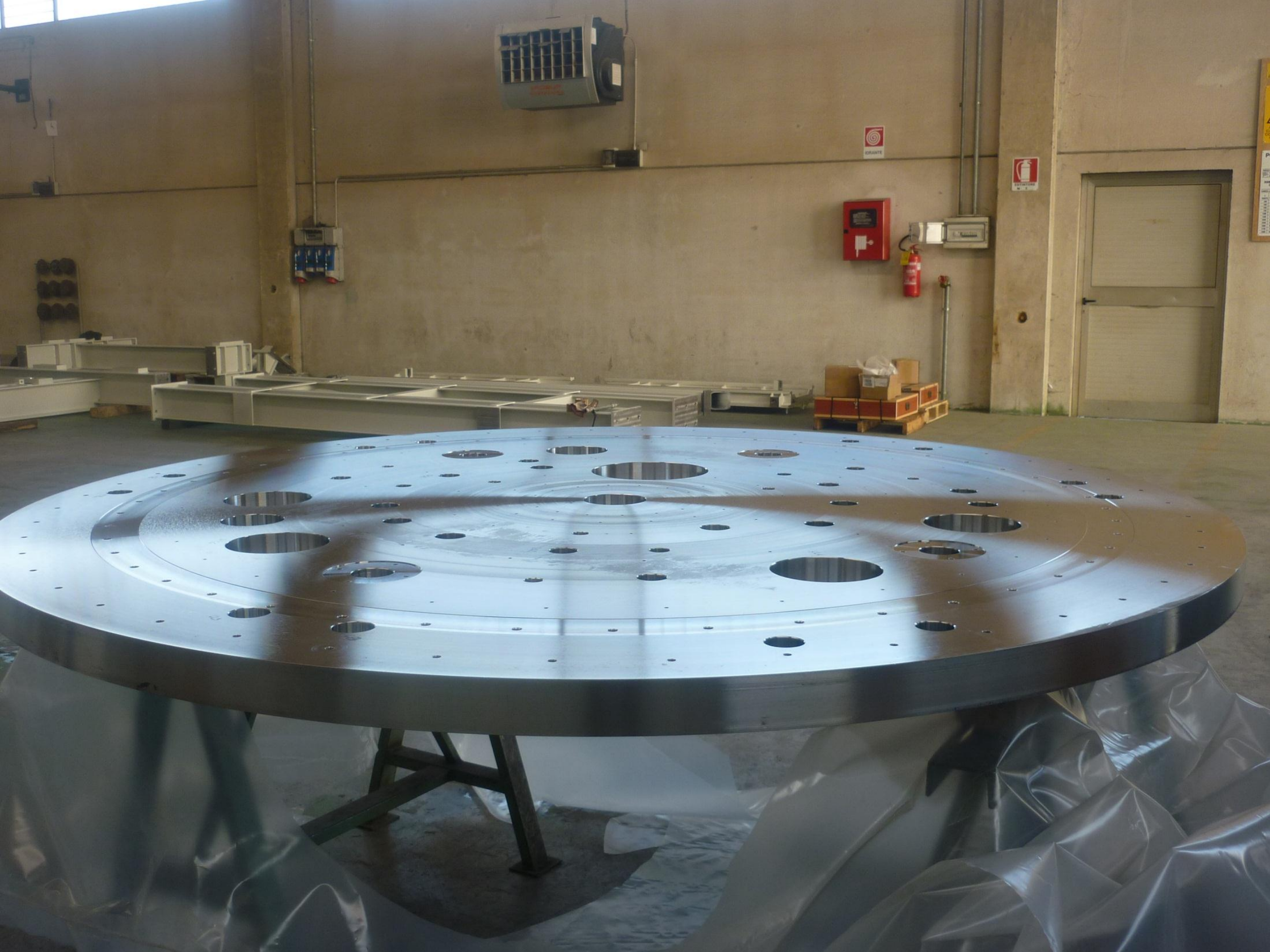} 
\caption{The 300~K plate after cleaning at the production site.}
\label{fig:cleaning_300k}
\end{figure}

The surface contamination of the vessels was mainly due to machine oil
deposited during calendering
(Fig.\ref{fig:cleaning_4k}). In most cases, pre-processing
was necessary to remove the bulk of the machine oil. Such
pre-processing was done with degreaser (10\% weight), nitric acid
($\sim$1\% weight) in hot ($\sim 40^\circ$C) unpurified water
solution. All gilded parts were protected by a peelable rubber coating 
(Green Maskant), which was removed before the installation at the production site (see
Sec.~\ref{sec:test_setup}). The 300\,K plate, which is made from AISI
304L steel, was cleaned by INFN personnel at the production site.  In this case, only the
degreasing cycles and the final flushing with Grade 2 water were
performed (Fig.\ref{fig:cleaning_300k}).

The effectiveness of this procedure in terms of radiopurity will be
tested at LNGS during the commissioning of the apparatus while the
vacuum tests at the production site have shown that the degassing rate
is low enough to achieve the required vacuum level (see
Sec.~\ref{sec:vacuum_tests}).

  \section{Test setup}
\label{sec:test_setup}

The CUORE Collaboration and Simic s.p.a. have built a dedicated test
facility to assess the quality of the produced items before final
assembly at LNGS. The aims of the tests were the evaluation of the
vacuum, pressure and thermal requirements described in
Sec.~\ref{sec:outer_shell}. As mentioned above, thermal properties
were tested only at liquid nitrogen temperature, where most of the
thermal stresses have already occurred. No radiopurity tests were
performed in this site.

\begin{figure}[ht]
\centering
\includegraphics[width=0.5\textwidth,keepaspectratio=true]{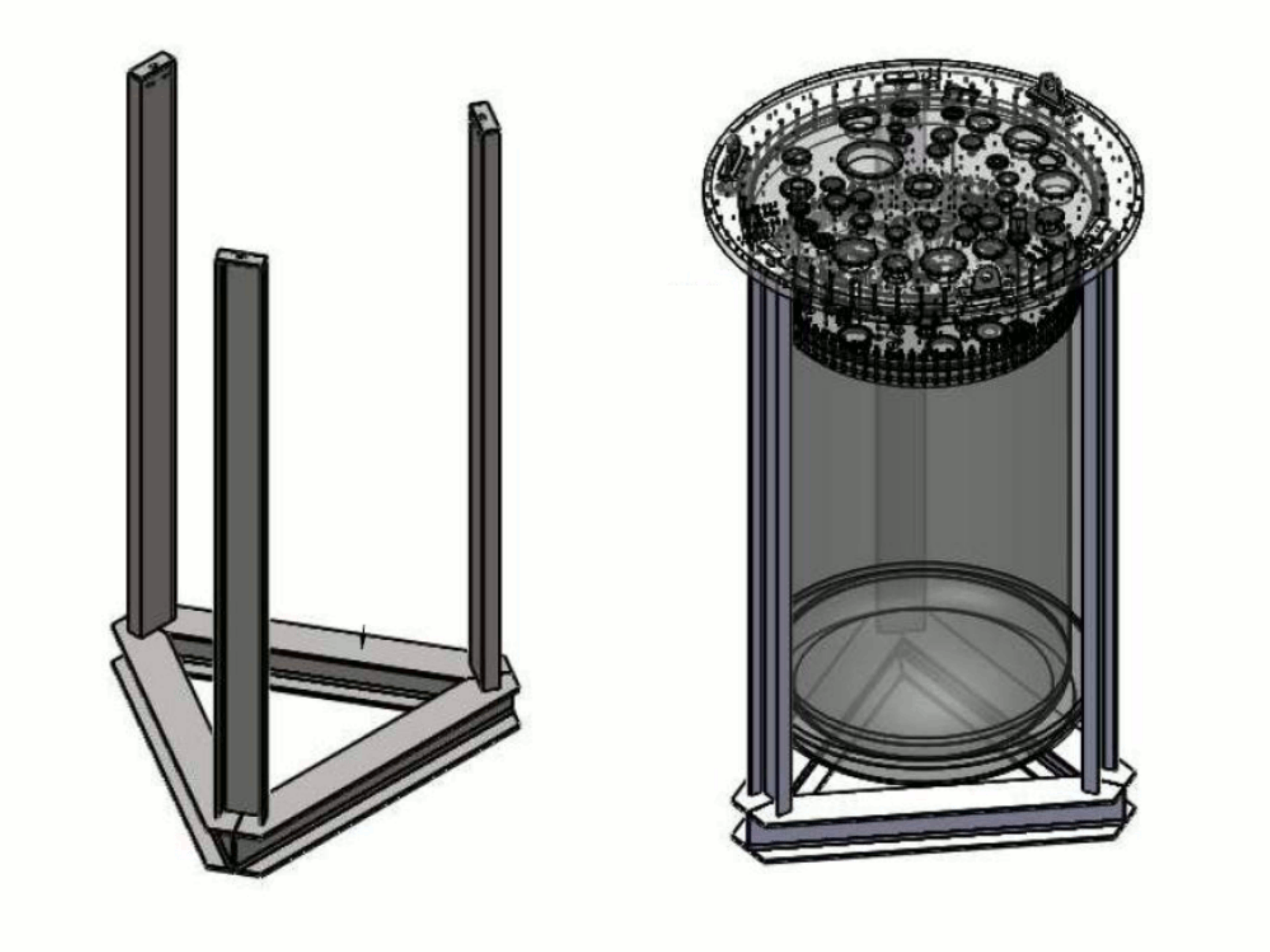} 
\caption{Iron support structure for the vessels at the production site (Courtesy of Simic s.p.a.).}
\label{fig:supporto}
\end{figure}

The test setup was built using the 300\,K and 4\,K vessels with their
plates. The 300\,K plate was connected to the 4\,K plate using the
A-bars, as in the final CUORE setup, while the 40\,K vessel and plate
were missing.  The two plates were first connected together by the
A-bars with the selected ports in place, then they were lowered onto
the 4\,K vessel. The vessel was sealed with the Helicoflex and, later
on, the 300\,K plate and the 4K plate-vessel assembly were positioned
on top of the 300\,K vessel. This setup was held in vertical position
by an iron support structure located near the pumps and the cooling
system (see Fig.\,\ref{fig:supporto}).  Only a few selected ports (see
Sec.~\ref{sec:outer_shell}) were employed during the tests, the 300\,K
and 4\,K opening of the others were sealed with blanks. The use of the
ports varied during the tests as described in Secs.~\ref{sec:results}.
Before closing the 4\,K chamber, each port was individually tested for
helium tightness of all sealings to a level of about
$10^{-9}\mathrm{mbar}\cdot\mathrm{l/s}$.  High vacuum (down to
$10^{-6}$~mbar at room temperature) was achieved using a
0.017\,m$^3$/s rotary pump coupled with a 0.3\,m$^3$/s hybrid
turbo-molecular pump (ATH300, Adixen). The turbo-molecular pump was
mounted just on the top of the IVC or OVC ports, to reduce flow
impedance. The rotary pump was located on the ground floor and
connected to the turbo through a DN25 flexible line. The cooling
system was based on regulated flux of liquid Nitrogen (LN$_2$) stored
in a 3000~l dewar and connected by a flexible line to the input port
of the fast cooling system (FCS). A second port was employed as 
exhaust to keep the IVC at atmospheric pressure during cool-down. The
LN$_2$ flow was regulated by a cryogenic solenoid valve positioned on
the output line of the dewar.  The cryogenic valve was driven by a set
of thermometers located inside the test setup as described in
Sec.~\ref{sec:cold_tests}.

  \section{Quality control tests}
\label{sec:results}


  \subsection{Vacuum tests}
\label{sec:vacuum_tests}

Vacuum tests at room temperature were performed both for the IVC and
the OVC. To test individually each welding of the 4\,K vessel, part of
the tests were performed before the insertion inside the 300~K vessel.
The 4\,K vessel was, hence, evacuated until it reached a stable
pressure of $6\times 10^{-6}$~mbar in 15 hours. A leak detector was
positioned on the back of the turbo pump in replacement of the rotary
pump. The sensitivity reached by the detector (background leak) was
$<4 \times 10^{-10} \mathrm{mbar}\cdot\mathrm{l/s}$. Gas tightness of
the welding was tested using plastic films stuck to the vessel along
the welding seams using adhesive tape to form bags. Helium was
injected into the bags and the growth of the background leak was
measured for no less than 2 minutes. No leak signal over the
background was observed for any of the 4~K vessel weldings. The same tests
were performed after the pressure tests and two thermal cycles. In
this case the background leak was significantly larger ($4 \times
10^{-9} \mathrm{mbar}\cdot\mathrm{l/s}$) because the IVC was previously flushed
with a He (50\%) - Ar (50\%) gas mixture during the pressure tests
(see below). Again, no leaks were observed for any of the
weldings. Similarly, a vacuum test of the OVC was performed using an
independent vacuum system with the turbo pump located near the rotary
pump. Due to the impedance between the pump and the OVC port, the
pressure reached by this system was $6 \times 10^{-4}$~mbar. 
Each welding was tested individually and no leak above the background
($1.3 \times 10^{-9} \mathrm{mbar}\cdot\mathrm{l/s}$) was observed.
The same test was performed after the pressure test of the OVC. This
time the turbo pump was located near the OVC port and a vacuum of $8
\times 10^{-5}$~mbar was reached in 15 hours; the leak background was
$6 \times 10^{-9} \mathrm{mbar}\cdot\mathrm{l/s}$ and no increase of
the leak rate was observed during the He tests with plastic bags.

The quality of the seals was also assessed.  The sealing of the
metallic gasket (Garlock Helicoflex\textsuperscript{\textregistered},
model HNV 200) was checked with special care due to the unusual
configuration employed in CUORE. The gasket is positioned on top of
the 4\,K flange, on a 0.5~mm relief machined in the flange of the
vessel. The measured surface roughness of the relief is 0.78~$\mu$m
RA~\cite{ISO_roughness}. The gasket is fixed by a centering ring
running outside the Helicoflex (height $5.7 \pm 0.1$~mm). Another
0.5~mm relief (surface roughness 0.59~$\mu$m RA) is machined in the
4\,K plate and during the sealing of the vessel, the 6.9~mm thick
gasket is pressed and reaches a height of 5.7~mm. The 4\,K plate is
fixed on the vessel with 80 M12 bolts tensioned with nuts and equipped
with washers made of Ti-6Al-4V ELI alloy to compensate for the
difference in the steel and copper thermal contraction. The bolts were
tensioned using three hydraulic tensioners positioned $120^\circ$
apart. Three rounds of tensioning were performed to reach uniformly
the nominal distance between the 4\,K vessel and plate. The bolts were
tensioned to about 60\,bar, which corresponds to a load of about
10\,kN.

The Helicoflex showed no evidence of leaks at the level of $4 \times
10^{-10} \mathrm{mbar}\cdot\mathrm{l/s}$ (background of the IVC vacuum
test).  Some problems were experienced, however, with the indium
flanges of the IVC central port and the FCS ports and for the s-shaped
bellows of the FCS.  The former was traced back to a mechanical
mismatch between the steel flange of the bellows and the design of the
copper flange, resulting in a larger gap for the indium.  It was cured
increasing from 1 to 1.5~mm the diameter of the indium wire. The
defective s-shaped bellows was due to porosities in the bulk
material, which was temporary fixed (at room temperature) using a
vacuum plasticine. After these treatments, vacuum tightness was
recovered at the $4 \times 10^{-9} \mathrm{mbar}\cdot\mathrm{l/s}$
level.

  \subsection{Pressure tests}
\label{sec:pressure_tests}

Pressure tests are aimed at inducing mechanical stresses on the
vessels and evaluating the occurrence of permanent deformations
or worsening of the vacuum tightness at the weldings. They also conservatively test 
the design specifications of Tab~\ref{tab:pressure}. Since during
thermal cycles the IVC undergoes further stresses, the pressure test
was carried out for the IVC before and after the cold tests. In
particular, we performed the following tests:

\begin{itemize}
\item over-pressuring the OVC by 0.3 barg, i.e. by a relative
  pressure of 30~kPa (absolute pressure: $p_{OVC}=$130~kPa = 1.3~bara) while
  pumping the IVC ($p_{IVC}\simeq 0$~bara)
\item over-pressuring the IVC by 0.5 barg ($p_{IVC}=1.5$~bara) while
  pumping the OVC ($p_{OVC}\simeq 0$~bara)
\item over-pressuring both the IVC and the OVC by 0.5 barg
  ($p_{IVC}=p_{OVC}=1.5$~bara)
\end{itemize}

\begin{figure}
\centering
\includegraphics[scale=0.23,type=pdf,ext=.pdf,read=.pdf]{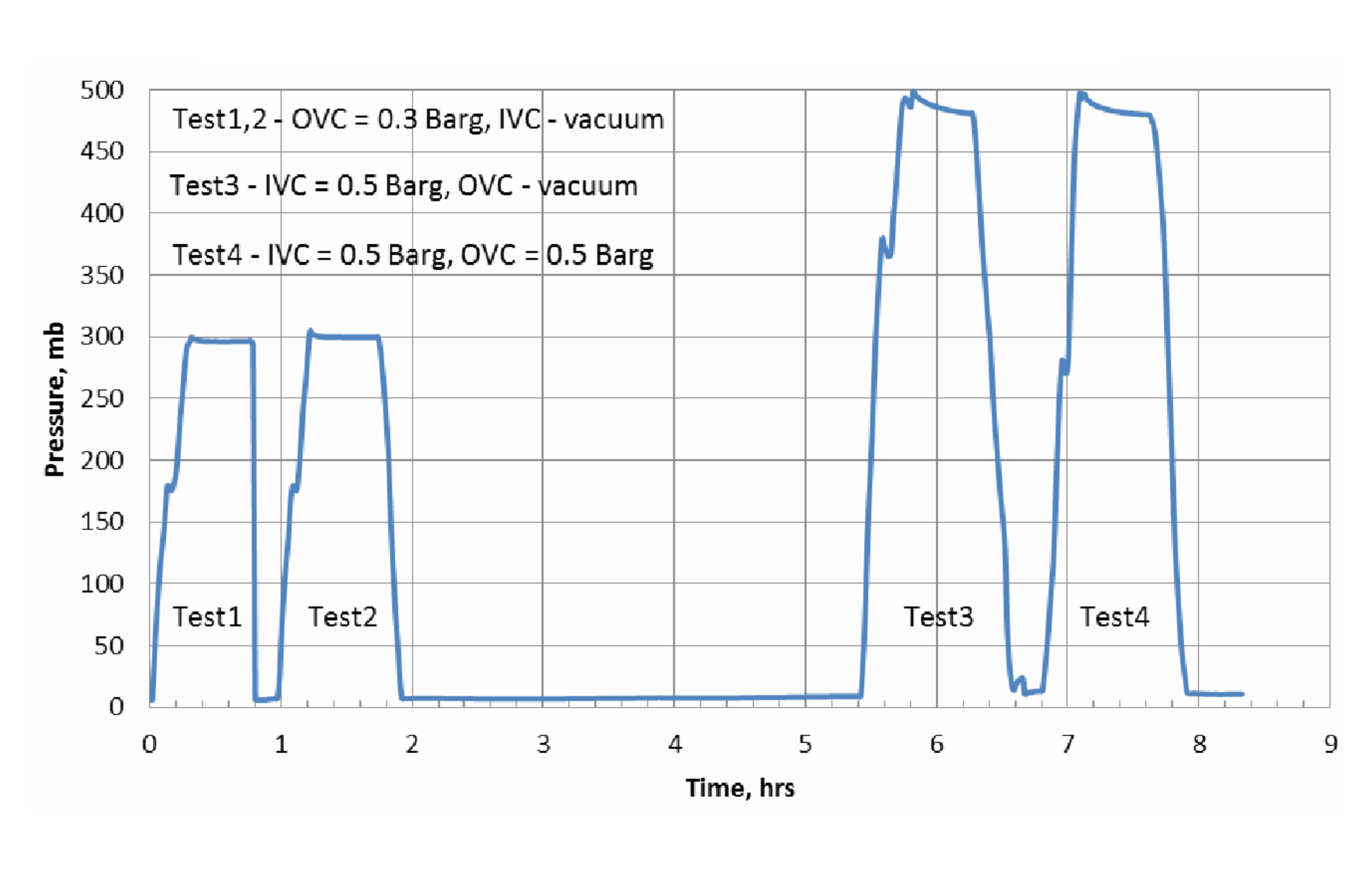} 
\caption{Pressure versus time in the IVC and OVC.}
\label{fig:pressure_tests}
\end{figure}

Overpressure was obtained injecting a He (50\%) -Ar (50\%) gas mixture
in the IVC (OVC) port, while pumping the OVC (IVC). For greater
safety, the test was performed filling a buffer bottle from a high
pressure reservoir. The pressure of the buffer bottle ($p_{b}$) was
chosen so that the expansion of the gas into the IVC (or OVC) was enough to
bring the IVC to the sought-for pressure ($p_{IVC} = p_b
V_b/V_{IVC}$, $V_b$ and $V_{IVC}$ being the volume of the buffer
bottle and IVC respectively). In this way, there is no risk of
accidental overpressure in the vessels even in case of a malfunction
of the pressure reducer at the reservoir. During the expansion, the
pressure at IVC/OVC was monitored by a digital pressure gauge
and recorded by the DAQ system. The overpressure was kept for $\sim$30~min
in each cycle and every cycle was repeated twice (see Fig.~\ref{fig:pressure_tests}).

No visible deformation was observed during any of the cycles and, as
mentioned in Sec.~\ref{sec:vacuum_tests}, vacuum tightness at the
weldings was preserved. In addition, a dedicated campaign was taken to
check microdeformations of the IVC at the end of the thermal and
pressure tests.  The measurements were taken using a Laser Tracker and
compared with the data recorded at the end of the vessel
production (Nov 2011). The maximum difference with respect to previous
measurements was 0.1~mm for the diameter and 0.6~mm for the roundness.

  \subsection{Cold tests}
\label{sec:cold_tests}

Two thermal cycles were performed bringing the IVC to about 82~K and
back to room temperature. Thermal cycles induce stresses in the
weldings due to copper contractions at cool-down and expansions at
warm-up. Since more than 90\% of the linear thermal contraction of copper\footnote{For Cu~\cite{ekin},
  the total linear contraction $\Delta L/L \equiv
  (L_{293K}-L_T)/L_{293K}$ is $3.24 \times 10^{-3}$ at $T=4$K and
  $3.02 \times 10^{-3}$ at $T=77$K.} already occurs at 77~K, cooling
with LN$_2$ is adequate to  test the outer cryostat of
CUORE.

\begin{figure}
\centering
\includegraphics[scale=0.37,type=pdf,ext=.pdf,read=.pdf]{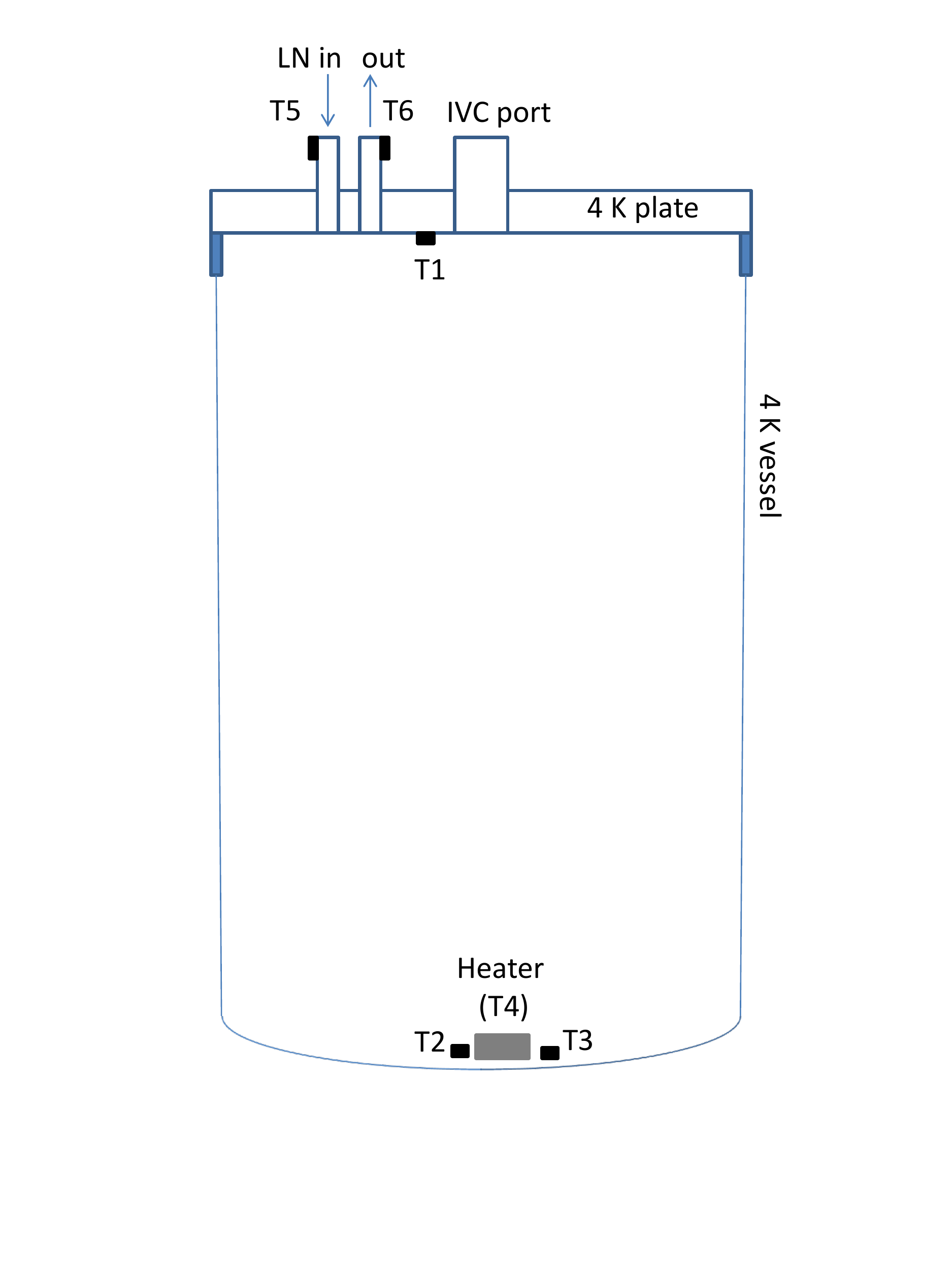} 
\caption{Position of the thermometers T1-T6.}
\label{fig:thermometers}
\end{figure}

As anticipated in Sec.~\ref{sec:test_setup}, cooling was achieved
circulating LN$_2$ from the input port of the FCS. The Nitrogen was
extracted by the dewar at a pressure of 2~bar at the entrance of the
cryovalve. The opening of the cryovalve was driven by four PT100
thermometers (T1, T2, T5, T7 in Fig.~\ref{fig:thermometers}). T1 and
T2 were located at the bottom and top of the 4\,K vessel,
respectively. T5 and T7 at the end and at the beginning of the line
connecting the cryovalve to the FCS port. The cryovalve was
opened/closed through a relay by a Lake Shore model 218 Temperature
Monitor. The valve was open when the following conditions were
fulfilled: T2$>$80~K (no LN$_2$ accumulating in the bottom of the vessel),
T1-T2$<$30~K (temperature gradient along the height of the vessel
$<$30~K), T5-T7$<$200~K (a safety condition to test the integrity of
the transfer line). In addition, the temperature of the Nitrogen vapor
at the exit of the vessel, i.e. at the output port of the FCS, was
monitored (T6). A 60~W heater was also installed at the bottom of the
vessel to speed up the warming. The regulation system, however,
operated smoothly down to the target temperature (T1=82~K) and the
heater was not used in any phase of the cycle. Finally, T3 and T4 were
employed as spare for T2 and as T monitor in the proximity of the heater,
respectively.

\begin{figure}[ht]
\centering
\includegraphics[scale=0.2,type=pdf,ext=.pdf,read=.pdf]{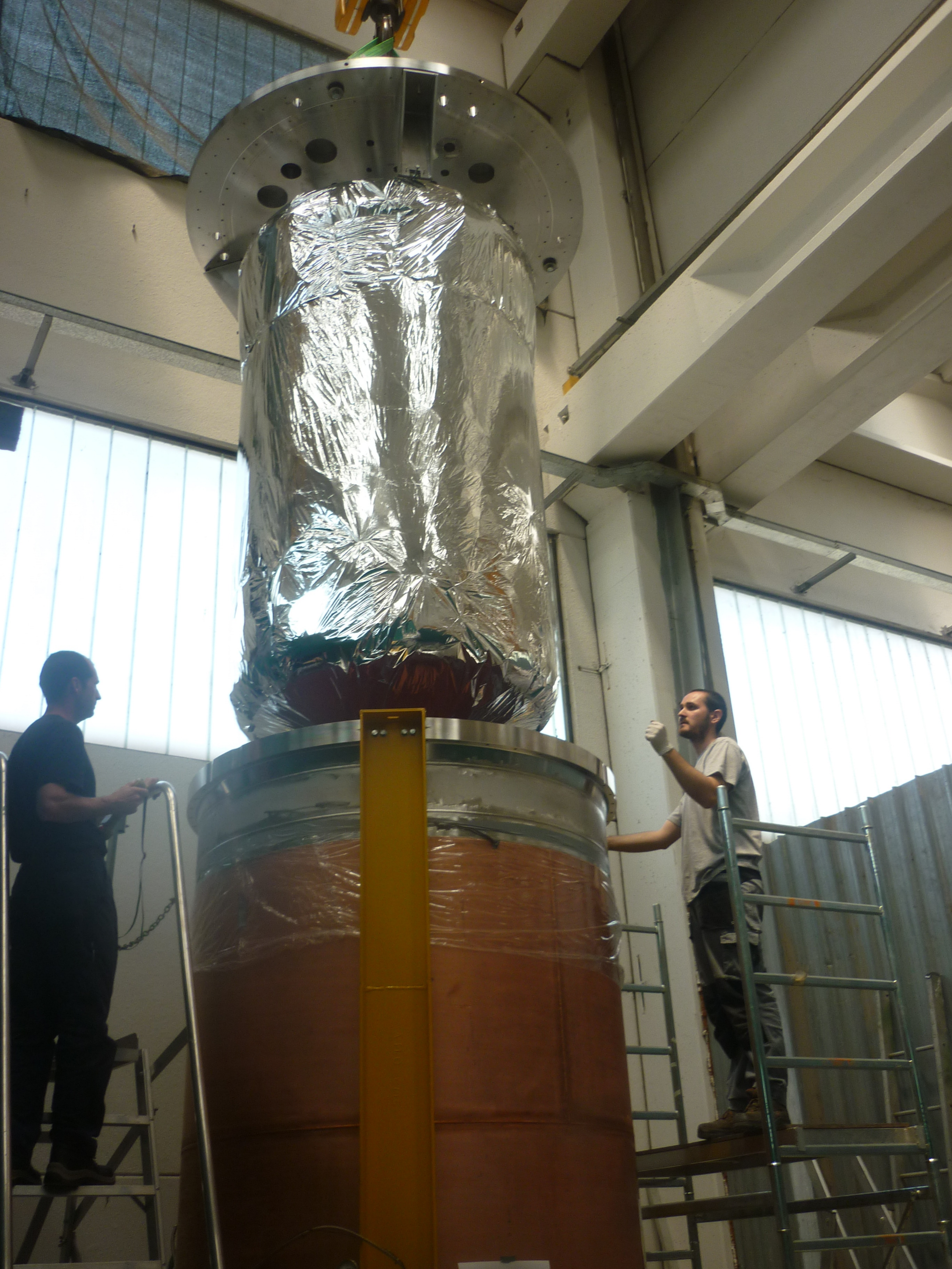} 
\caption{Insertion of the IVC into the 300\,K vessel before the cold
  tests. The 4\,K vessel and plate is wrapped with superinsulation.}
\label{fig:superinsulation}
\end{figure}

Before the cold test, the 4\,K vessel and plate were wrapped with a 10 layer
aluminized superinsulation blanket (Ruag Space GmbH, model COOLCAT II, 
see Fig.~\ref{fig:superinsulation}). During cooldown, the IVC port was equipped with a
rupture disk
with burst pressure $0.45\pm 0.05$~bar, a drop-off plate (300~mbar)
and connected to the pressure gauge used during the tests of
Sec.~\ref{sec:pressure_tests}. Along this line, a pressure switch set
at 200~mbar was connected in parallel to the relay driving the
cryovalve.  Hence, the maximum overpressure causing a stop of the
cooling was set to 200~mbar while exhaust through the IVC port was
granted at 300~mbar. The OVC port was used to connect the turbo
pump to the OVC and record the pressure with a high-vacuum pressure
gauge. The wire port was employed for the readout of
the thermometers and to drive the heater with a 24V/3A DC current
generator.

\begin{figure}[ht]
\centering
\includegraphics[width=0.95\linewidth,type=pdf,ext=.pdf,read=.pdf]{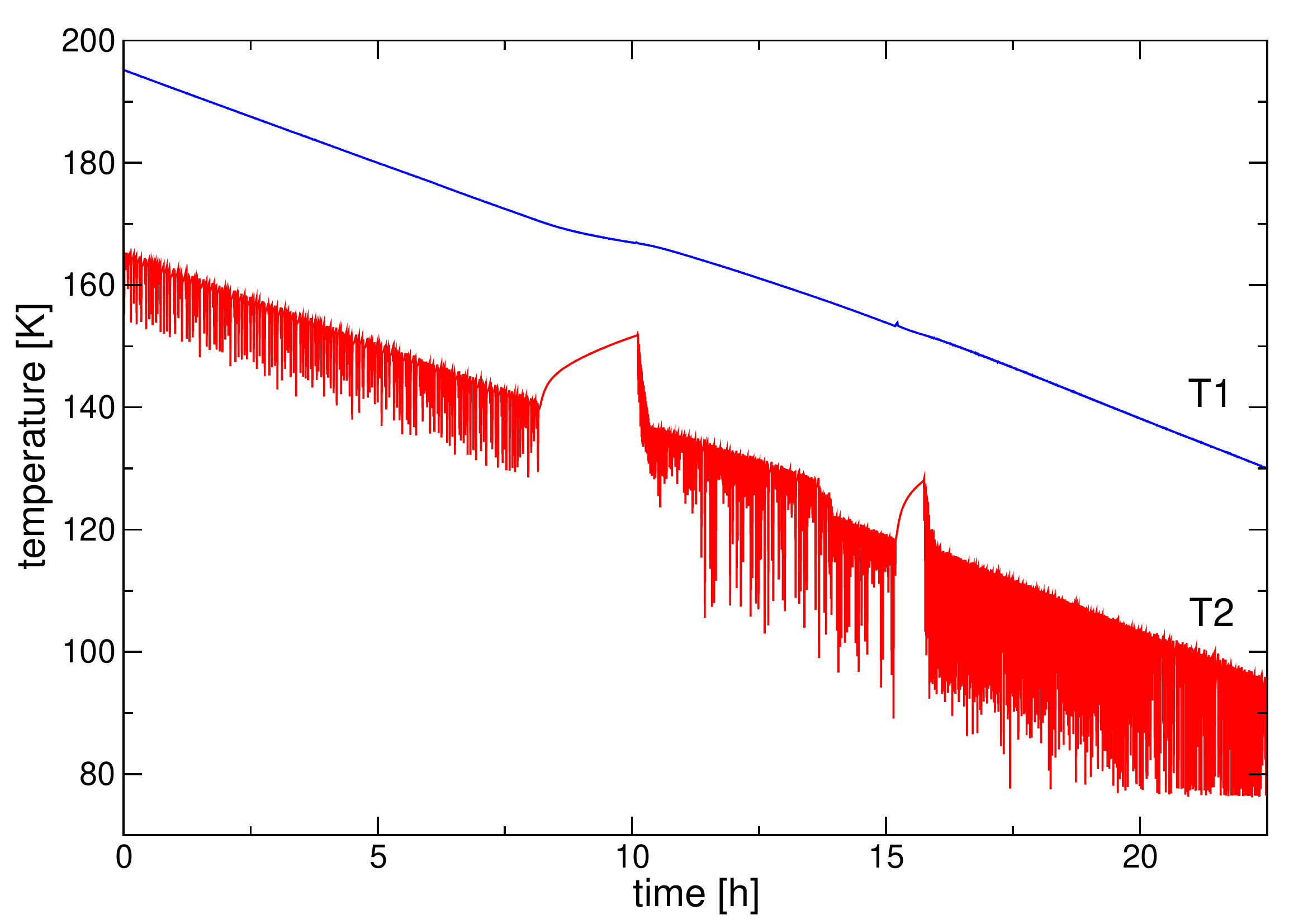} 
\caption{Temperature at T1 (4\,K plate) and T2 (bottom of the 4\,K vessel) during
cooldown. The temperature jumps at T2 occurred after 8 and 15 hours from the beginning
were caused by failure of the silicon tube section in the LN$_2$ transfer line.}
\label{fig:cooling}
\end{figure}

Cooldown started with a pressure in the OVC of $p_{OVC}=3.8 \times
10^{-5}$ and $1.8 \times 10^{-5}$~mbar in the first and second cycle,
respectively. During cooldown it reached a minimum value of $p_{OVC}=5
\times 10^{-8}$~mbar. The cooldown rate was stable (see
Fig.~\ref{fig:cooling}) at about 3~K/h and the opening of the cryovalve
was driven by the T1-T2$<$30~K condition until T1$\simeq 110$~K,
i.e. when the temperature at T2 was close to LN$_2$ temperature. Then,
cool-down was mostly driven by the T1$>$80~K condition. The
temperature reached at the end of the cooling was 82.5~K (84~K) at T1
and 77~K (77~K) at T2 during the first (second) cycle. The LN$_2$
consumption per cycle was about 1800~l, the total mass to be cooled
being 2012~kg. As expected, LN$_2$ consumption was $\sim 2$ times larger
than for an ideal cooling system that uses only the latent heat of
vaporization (900~l~\cite{ekin}).  Warm-up was purely passive (no use
of heaters) in both cycles. While keeping high vacuum in the OVC (
$p_{OVC}=5 \times 10^{-3}$~mbar with the turbo pump off) the warming
rate was about 3~K/h and it reached 4~K/h when $p_{OVC}$ was brought
to 1~mbar. In the final phase of the warm-up (T1$\simeq$260~K), in
order to increase the heat exchange and maintain a warming rate of
4~K/h, $p_{OVC}$ was brought to 50~mbar.

In both cycles, a leak was observed at low temperatures. The leak
developed at 170~K (160~K) at the first (second) cycle and worsened
the OVC vacuum up to $p_{OVC}=1 \times 10^{-4}$~mbar at the end of the
cooling. The integral leak rate was measured injecting the He-Ar
mixture in the IVC (see Sec.~\ref{sec:pressure_tests}) and measuring
the He flow in the OVC with the leak detector positioned at the back
of the turbo pump; at 77 K the leak amounted to $3 \times 10^{-4}
\mathrm{mbar}\cdot\mathrm{l/s}$. During the warm-up the integral leak
rate decreased but was still visible at room temperature ($1.4 \times
10^{-7} \mathrm{mbar}\cdot\mathrm{l/s}$). Finally, in both cycles the
leak ($1 \times 10^{-7} \mathrm{mbar}\cdot\mathrm{l/s}$) was located
at the FCS indium flanges and s-tubes using He injected into plastic
bags. No additional leaks were observed in any other port, in the
Helicoflex and in the weldings.  After the first cycle, the leak at
the FCS was temporary fixed tightening the screws of the indium flanges
but due to the mechanical defects of the FCS tubes
(Sec.\ref{sec:vacuum_tests}) a full repair was not possible. The
FCS ports are therefore the most likely explanation of the
deterioration of $p_{OVC}$ observed during the cooling.

\subsection{Final tests and delivery to LNGS}
\label{sec:delivery}
As mentioned above (see Sec.~\ref{sec:pressure_tests}), after the
second thermal cycle and warm-up, a pressure test was performed on the
IVC ($p_{IVC}=1.5$~bara, $p_{OVC} = 4 \times 10^{-2}$~mbar), followed
by a leak test of the weldings and the measurement campaign performed
with the laser tracker.  The system was hence dismounted and packed
under Nitrogen atmosphere using the same packing selected for BAMA s.r.l..
Finally, a mechanical test, i.e. a mounting test of the plate on its
vessel, was performed for the 40~K shield, which does not require
vacuum tightness. The vessels and plates were delivered to LNGS on
July 19 2012.

  \section{Conclusions}
\label{sec:conclusions}

In this paper, we described the construction and the validation tests
of the outer cryostat for the CUORE experiment. In particular, we have
demostrated that the electron-beam welding (EBW) is an affordable
technique for the assembly of large copper cryostats when radioactive
contaminants from filler metals or electrodes must be avoided. The
quality of the EBW welding has been tested with a measurement campaign
carried out at the production site (Camerana, Italy). The campaign
included mechanical and vacuum tightness tests of the weldings and of
the whole vessels. Vacuum tightness has been demonstrated at the level
of $10^{-9} \mathrm{mbar}\cdot\mathrm{l/s}$ both for the copper-steel
Outer Vacuum Chamber (OVC) and for the all-copper Inner Vacuum Chamber
(IVC) of CUORE. The IVC and OVC were mechanically stressed by
overpressure cycles at 1.3 and 1.5 bara and by two thermal cycles at
82~K. The vessels withstood the overpressure and underpressure
conditions without structural collapse or permanent deformations. In
particular, no deformations of the IVC larger than 0.6~mm were
observed after pressure and thermal cycles. Similarly, vacuum
tightness of all components was preserved during the tests with
maximum leak rate less than $10^{-9}
\mathrm{mbar}\cdot\mathrm{l/s}$. The only anomaly reported is a vacuum
leak that developed at T$<$170~K; this anomaly has been traced back to
the defective fast cooling tubes for the IVC. Finally, the A-bars and
the 4~K metallic seal (Helicoflex) have been tested in the same
conditions as for CUORE and were shown to be compliant with
specifications.

The 4\,K outer cryostat of the CUORE experiment has been delivered to the
Gran Sasso Underground Laboratories in July 2012 and  is currently
under commissioning.

  \section*{Acknowledgments}
We wish to express our gratitude to Simic s.p.a. for support during
the tests at the production site. We are indebted to C. Bucci,
A.~Franceschi, S.~Gazzana, P.~Gorla, K.~Heeger, R.~Kadel,
T.~Napolitano, A.~Pelosi, C.~Zarra and S.~Zucchelli for many useful
remarks about the validation tests. We are also grateful to
S.~Castoldi and G.~De Bernardi (BAMA s.r.l.) for suggestions during
the setting up of the cleaning procedures. Finally, we thank our colleagues 
of the CUORE Collaboration and, in particular, C.~Brofferio, A.~Dally, K.~Heeger, 
S.~Sangiorgio and N.~Scielzo for their careful reading of the manuscript.

\end{document}